\documentclass[twocolumn]{autart}    

\usepackage{tikz}
\usepackage{amsmath}
\usepackage{latexsym, amssymb, amsmath, amsbsy,amsopn, amstext}
\usepackage{graphicx,hyperref,booktabs}
\usepackage{cancel, color}
\usepackage{epstopdf,subcaption}
\graphicspath{{images/}}

\DeclareMathOperator{\Row}{Row}

\def\cal{\mathcal}

\def\ra{\rightarrow}

\def\lra{\leftrightarrow}

\def\a{\alpha}
\def\b{\beta}
\def\d{\delta}

\def\D{\Delta}

\def\L{\Lambda}

\def\0{{\bf 0}}

\def\con{con}
\def\J{{\bf 1}}

\newcommand{\B}{{\mathcal B}}

\def\dsum{\mathop{\sum}\limits}
\newtheorem{dfn}[thm]{Definition}
\newtheorem{prp}[thm]{Proposition}
\newtheorem{exa}[thm]{Example}

\begin{document}

\begin{frontmatter}

\title{Aggregated (Bi-)Simulation of Finite Valued Networks \thanksref{footnoteinfo}}
\thanks[footnoteinfo]{This work is supported partly by NNSF
    62073315 of China. Corresponding author: Daizhan Cheng. Tel.: +86 10 6265 1445; fax.: +86 10 6258 7343.}

\author{Zhengping Ji\dag\ddag, Xiao Zhang\dag, Daizhan Cheng\dag}

\address{\dag Key Laboratory of Systems and Control, Institute of Systems Science,\\
Chinese Academy of Sciences, Beijing 100190, P.R.China}
\address{\ddag School of Mathematical Sciences, University of Chinese Academy of Sciences, Beijing 100049, P.R.China}


\begin{abstract}
The paper provides a method to approximate a large-scale finite-valued network by a smaller model called the aggregated simulation, which is a combination of aggregation and (bi-)simulation. First, the algebraic state space representation (ASSR) of a transition system is presented. Under output equivalence, the quotient system is obtained, which is called the simulation of the original transition system. The ASSR of the quotient system is obtained. The aggregated (bi-)simulation is execueted in several steps: a large scale finite-valued network is firstly aggregated into several blocks, each of which is considered as a network where the in-degree nodes and out-degree nodes are considered as the block inputs and block outputs respectively. Then the dynamics of each block is converted into its quotient system, called its simulation. Then the overall network can be approximated by the quotient systems of each blocks, which is called the aggregated simulation. If the simulation of a block is a bi-simulation, the approximation becomes a lossless transformation. Otherwise, the quotient system is only a (non-deterministic) transition system, and it can be replaced by a probabilistic networks. Aggregated simulation can reduce the dimension of the original network, while a tradeoff between computation complexity and approximation error need to be decided.
\end{abstract}

\end{frontmatter}

\section{Introduction}

In recent years networked systems, as one of the most important objects of complex systems,  become a hot topic in overall scientific community.
There are so many networks from high technologies to daily life, such as internet, neural network, logistic network, Boolean network for gen regularity network, networked game, etc.

Among them, finite-valued network is one of the most important networks. The historical origin of finite-valued networks may be tracked to
finite-valued machine such as Turing machine \cite{tur50,yan22,yue21}; coding-decoding in cryptography \cite{ber68,car10,zho21};
mapping and functions over finite sets \cite{bir67,rob86,sch76}; and Boolean networks \cite{che11,hua00,kau69,kau93,kau95}; etc. In the early 1960s
two Robel price winners Jacob and Monod showed that any cell contains a number of regulatory genes that acts as switches, which inspired Kauffman to propose a model, called Boolean network, to formulate genetic regulatory network \cite{wal92}. Since the Boolean network has been successfully applied to biological networks, it caused a enthusiasm in studying Boolean network and Boolean control network.

 A Boolean network is described as a logical system, there was no convenient tool to deal with it, until semi-tensor product (STP) of matrix was used into the investigation. STP converts a logical (control) system into a linear (bilinear) difference system, called the algebraic state space representation (ASSR). Then the classical mathematical tools, such as the matrix theory and the theory of difference equation etc., can be used to analysing and control designing of Boolean networks \cite{che11}.  This approach promotes the investigation of Boolean networks as well as other kind of finite valued networks.

Nowadays, the classes of finite valued networks investigated are various. According to the domain values there are Boolean networks, $k$-valued networks \cite{ada03}, and mix-valued networks \cite{che11}; according to the state transition form, there are deterministic (which means there is at most one successor to each state such as Boolean networks) and non-deterministic (which means a state may have more than one successors such as finite automata), and the transition type maybe conventional (i.e., uniquely determined), probabilistic \cite{shm02}, and stochastic \cite{che13}; according to the algebraic structure of the bearing space, there are Boolean algebra \cite{che12}, finite field \cite{men20}, finite ring \cite{che22}, lattice \cite{jipr}, etc. In addition, several other kinds of networks can be converted into such category, such as finite networked games \cite{che21}, finite valued machines \cite{yan22,yue21}, etc.

All the above mentioned different kinds of finite valued networks have been investigated using STP. Various of analysis and control problems have been discussed, including topological structure \cite{che10}, controllability, observability, disturbance decoupling, decomposition, identification, realization, tracking, and optimal control, etc. We refer to some survey papers for the application of STP to control problems of finite-valued networks \cite{for16,muh16,lu17,li18}.

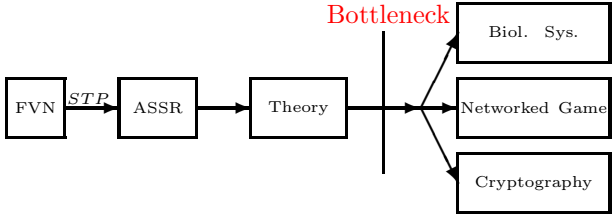
\begin{figure}
\centering
\setlength{\unitlength}{0.5cm}
\begin{picture}(16,6)\thicklines
\put(0,2){\framebox(1.5,1.5){\begin{tiny}FVN\end{tiny}}}
\put(3,2){\framebox(2,1.5){\begin{tiny}ASSR\end{tiny}}}
\put(6.5,2){\framebox(2.5,1.5){\begin{tiny}Theory\end{tiny}}}
\put(12,0){\framebox(4,1.5){\begin{tiny}Cryptography\end{tiny}}}
\put(12,2){\framebox(4,1.5){\begin{tiny}Networked~Game\end{tiny}}}
\put(12,4){\framebox(4,1.5){\begin{tiny}Biol. ~Sys.\end{tiny}}}
\put(1.5,2.75){\vector(1,0){1.5}}
\put(5,2.75){\vector(1,0){1.5}}
\put(9,2.75){\vector(1,0){2}}
\put(11,2.75){\vector(1,2){1}}
\put(11,2.75){\vector(1,0){1}}
\put(11,2.75){\vector(1,-2){1}}
\put(1.6,2.85){\begin{tiny}$STP$\end{tiny}}
\put(8.5,5){{\color{red} Bottleneck}}
\thinlines
\put(10,1){\line(0,1){3.8}}
\end{picture}
\caption{STP Approach to FVN \label{Fig.agg.1.2}}
\end{figure}

As depicted in Fig. \ref{Fig.agg.1.2}, to apply STP approach to practical problems there is a bottleneck, that is, the computational complexity.

%
%

It is a common phenomenon that a large scale networked system may have nodes $|N|>>1$, but the degrees of nodes (in-degree and out-degree) are very small. Say, it was pointed by Kauffman \cite{kau95} that a genetic regularity network may have over a thousand of nodes but the degrees are always less than $10$. This fact makes an aggregation efficient in reducing the computational complexity.

\section{Transition Systems}

\begin{dfn}\label{d2.1} \cite{bel17} A tuple $T=(X,U,\Sigma,O,h)$ is called a transition system, where
\begin{itemize}
\item[(i)] $X$ is the set of states,
\item[(ii)] $U$ is the set of inputs (controls or actions),
\item[(iii)] $\Sigma:X\times U \ra 2^X$ is a transition mapping,
\item[(iv)] $O$ is the observations,
\item[(v)] $h:X\ra O$: observation mapping.
\end{itemize}
If $|\Sigma(x,u)|\leq 1$, $T$ is said to be deterministic.
\end{dfn}

A conventional way to describe a transition system is using a transition graph. We use an example to explain this.

\begin{exa}\label{e2.2} Consider a transition system
$$T=(X,U,\Sigma,O,h),
$$ where
\begin{itemize}
\item[(i)] $X=\{x_1,x_2,x_3,x_4\}$,
\item[(ii)] $U=\{u_1,u_2\}$,
\item[(iii)]
$$
\begin{array}{ll}
\Sigma(x_1,u_1)=\{x_2,x_3\}&\Sigma(x_2,u_1)=\{x_2,x_3\}\\
\Sigma(x_2,u_2)=\{x_4\},&\Sigma(x_3,\sigma_2)=\{x_2,x_3\}\\
\Sigma(x_4,\sigma_1)=\{x_2,x_4\},&~\\
\end{array}
$$
\item[(iv)] $O=\{O_1,O_2,O_3\}$,
\item[(v)]
$$
h(x_1)=O_1,\quad h(x_2)=h(x_4)=O_2,\quad h(x_3)=O_3.
$$
\end{itemize}

The system can be described by Figure \ref{Fig.fm.1.1}).

\vskip 5mm

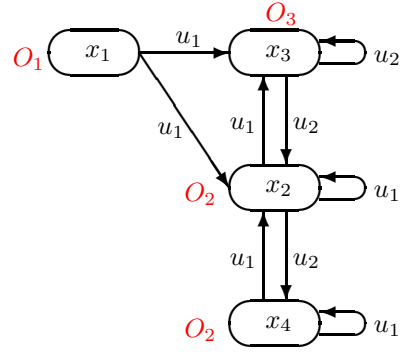
\begin{figure}
\centering
\setlength{\unitlength}{0.6cm}
\begin{picture}(9,9)\thicklines
\put(2,7.5){\oval(2,1)}
\put(6,1.5){\oval(2,1)}
\put(6,4.5){\oval(2,1)}
\put(6,7.5){\oval(2,1)}
\put(3,7.5){\vector(1,0){2}}
\put(3,7.5){\vector(2,-3){2}}
\put(5.75,5){\vector(0,1){2}}
\put(6.25,7){\vector(0,-1){2}}
\put(5.75,2){\vector(0,1){2}}
\put(6.25,4){\vector(0,-1){2}}
\put(7,1.5){\oval(2,0.5)[r]}
\put(7.5,1.75){\vector(-1,0){0.5}}
\put(7,4.5){\oval(2,0.5)[r]}
\put(7.5,4.75){\vector(-1,0){0.5}}
\put(7,7.5){\oval(2,0.5)[r]}
\put(7.5,7.75){\vector(-1,0){0.5}}
\put(1.8,7.4){$x_1$}
\put(5.8,7.4){$x_3$}
\put(5.8,4.4){$x_2$}
\put(5.8,1.4){$x_4$}
\put(3.8,7.7){$u_1$}
\put(3.4,5.7){$u_1$}
\put(5,5.8){$u_1$}
\put(5,2.8){$u_1$}
\put(6.4,5.8){$u_2$}
\put(6.4,2.8){$u_2$}
\put(8.2,7.3){$u_2$}
\put(8.2,4.3){$u_1$}
\put(8.2,1.3){$u_1$}
\put(0.2,7.2){${\color{red}O_1}$}
\put(5.8,8.2){${\color{red}O_3}$}
\put(4,4.2){${\color{red}O_2}$}
\put(4,1.2){${\color{red}O_2}$}
\end{picture}
\caption{Transition System in Example \ref{e2.2}\label{Fig.fm.1.1}}
\end{figure}

\end{exa}

We are only interested in finite transition system. Then similarly to multi-valued logic, a  finite transition system
has its  algebraic state space representation (ASSR).

\begin{prp}\label{p2.3} Consider a transition system $T=(X,U,\Sigma,O,h)$.
Assume ~$|X|=n$, $|U|=m$, $|O|=p$. Using vector form expression $X=\D_n$,  $U=\D_m$, $O=\D_p$, then
the $T$ has its ASSR as follows:
\begin{align}\label{2.1}
\begin{cases}
x(t+1)=Lx(t)u(t),\\
y(t)=Hx(t),\\
\end{cases}
\end{align}
where $L\in {\cal B}_{n\times nm}$ is a Boolean matrix, $H\in {\cal L}_{p\times n}$ is a logical matrix.
\end{prp}

\begin{exa}\label{e2.4} Recall Example \ref{e2.2}.
Set
$$
\begin{array}{ll}
x_i=\d_4^i,& i=1,2,3,4;\\
\sigma_j=\d_2^j,&j=1,2;\\
o_k=\d_3^k,& k=1,2,3.
\end{array}
$$
Then  $T$ has its ASSR as
\begin{align}\label{2.2}
\begin{cases}
x(t+1)=L\sigma(t)x(t),\\
y(t)=Hx(t),
\end{cases}
\end{align}
where,
$$
L=\begin{bmatrix}
0&0&0&0&0&0&0&0\\
1&1&0&1&0&0&1&0\\
1&1&0&0&0&0&1&0\\
0&0&0&1&0&1&0&0\\
\end{bmatrix}
$$
$$
H=\d_3[1,2,3,2].
$$
\end{exa}

%
%
%

\section{Quotient Systems and Simulation}

\begin{dfn}\label{d3.1} \cite{bel17} Consider a transition system $T=(X,U,\Sigma,O,h)$. Two states $x_1,x_2\in X$ are said to be observationally equivalent, denoted by $x_1\sim x_2$, if
$$
h(x_1)=h(x_2).
$$
The set of equivalence classes is denoted by $X/\sim$.
\end{dfn}

Let $Z\in X/\sim$ be an equivalent class, $\con(Z)\subset X$ is the set of $x\in Z$,
that is,
$$
\con(Z)=\{x\;|\;x\in Z\}\subset X,
$$
where $\con:X/\sim\ra 2^X$ is called the concretization mapping.

\begin{dfn}\label{d3.2} Consider a transition system $T=(X,U,\Sigma,O,h)$.  $T/\sim :=(X/\sim,U,\Sigma_{\sim},O,h_{\sim})$ is called the quotient system of $T$ under observability equivalence, where
\begin{itemize}
\item[(i)] $X/\sim=\{x/\sim\;|\; x\in X\}$ is the set of (observability) equivalence classes.
\item[(ii)] $U$ (original) set of inputs.
\item[(iii)] $\Sigma_{\sim}:X/\sim \times U\ra 2^{X/\sim}$ is defined as follows:
Assume $X_i,X_j\in X/\sim$. $X_j\in \Sigma_{\sim}(X_i,u)$, if and only if, there exist $x_i\in \con(X_i)$, $x_j\in \con(X_j)$ such that
$$
x_j\in \Sigma(x_i,u).
$$
\item[(iv)] $O$ (original) set of observations.
\item[(v)] $h_{\sim}:X/\sim \ra O$ is defined as follows:
$$
h_{\sim}(X_i):=h(x_i),\quad x_i\in \con(X_i).
$$
\end{itemize}
\end{dfn}

\begin{exa}\label{e3.3}
\begin{itemize}
\item[(i)] Recall Example \ref{e2.2}, where the transition system $T$ is described by Figure \ref{Fig.fm.1.1}). Under observability equivalence we have
$$
\con(X_1)=\{x_1\},~\con(X_2)=\{x_2,x_4\},~\con(X_3)=\{x_3\}.
$$
Hence, the quotient system  $T/\sim$ is described by Figure \ref{Fig.fm.5.1}.

\vskip 5mm

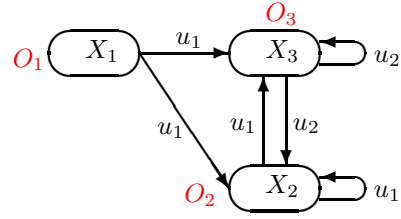
\begin{figure}
\centering
\setlength{\unitlength}{0.6cm}
\begin{picture}(9,6)(0,3.5)\thicklines
\put(2,7.5){\oval(2,1)}
\put(6,4.5){\oval(2,1)}
\put(6,7.5){\oval(2,1)}
\put(3,7.5){\vector(1,0){2}}
\put(3,7.5){\vector(2,-3){2}}
\put(5.75,5){\vector(0,1){2}}
\put(6.25,7){\vector(0,-1){2}}
%
\put(7,4.5){\oval(2,0.5)[r]}
\put(7.5,4.75){\vector(-1,0){0.5}}
\put(7,7.5){\oval(2,0.5)[r]}
\put(7.5,7.75){\vector(-1,0){0.5}}
\put(1.8,7.4){$X_1$}
\put(5.8,7.4){$X_3$}
\put(5.8,4.4){$X_2$}
\put(3.8,7.7){$u_1$}
\put(3.4,5.7){$u_1$}
\put(5,5.8){$u_1$}
\put(6.4,5.8){$u_2$}
\put(8.2,7.3){$u_2$}
\put(8.2,4.3){$u_1$}
\put(0.2,7.2){${\color{red}O_1}$}
\put(5.8,8.2){${\color{red}O_3}$}
\put(4,4.2){${\color{red}O_2}$}
\end{picture}
\caption{Quotient system of Example \ref{e2.2} (i) \label{Fig.fm.5.1}}
\end{figure}

\vskip 5mm

The ASSR of quotient system $T/\sim$ is easily obtained as
\begin{align}\label{3.1}
\begin{cases}
X(t+1)=L_q\sigma(t)X(t),\\
o(t)=H_qX(t),
\end{cases}
\end{align}
where,
$$
L_q=\begin{bmatrix}
0&0&0&0&0&0\\
1&1&0&0&1&1\\
1&1&0&0&0&1\\
\end{bmatrix}
$$
$$
H_q=I_3.
$$

\item[(ii)] Consider a transition system $T_2$ depicted by Figure \ref{Fig.fm.5.2}).

\vskip 5mm

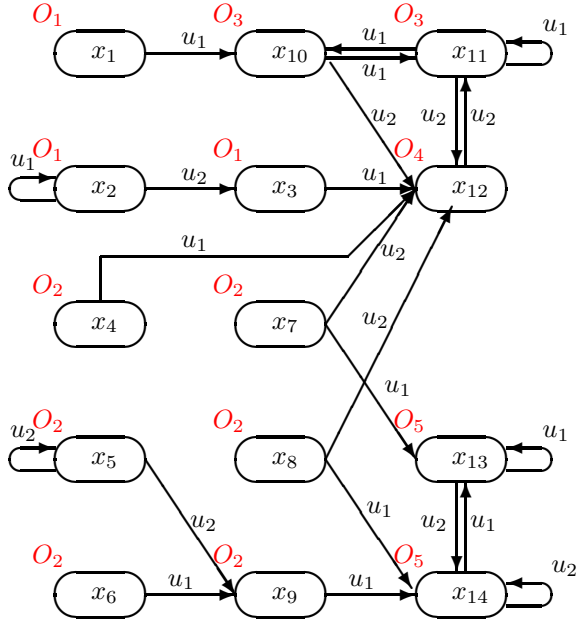
\begin{figure}
\centering
\setlength{\unitlength}{0.6cm}
\begin{picture}(13,15)\thicklines
\put(3,1.5){\oval(2,1)}
\put(3,4.5){\oval(2,1)}
\put(3,7.5){\oval(2,1)}
\put(3,10.5){\oval(2,1)}
\put(3,13.5){\oval(2,1)}
\put(7,1.5){\oval(2,1)}
\put(7,4.5){\oval(2,1)}
\put(7,7.5){\oval(2,1)}
\put(7,10.5){\oval(2,1)}
\put(7,13.5){\oval(2,1)}
\put(11,1.5){\oval(2,1)}
\put(11,4.5){\oval(2,1)}
\put(11,10.5){\oval(2,1)}
\put(11,13.5){\oval(2,1)}
\put(4,13.5){\vector(1,0){2}}
\put(8,13.4){\vector(1,0){2}}
\put(10,13.6){\vector(-1,0){2}}
\put(8.1,13.3){\vector(2,-3){1.9}}
\put(10.9,13){\vector(0,-1){2}}
\put(11.1,11){\vector(0,1){2}}
\put(8,10.5){\vector(1,0){2}}
\put(4,10.5){\vector(1,0){2}}
\put(3,8){\line(0,1){1}}
\put(3,9){\line(1,0){5.5}}
\put(8.5,9){\vector(1,1){1.5}}
\put(8,7.5){\vector(2,3){2}}
\put(8,7.5){\vector(2,-3){2}}
\put(4,4.5){\vector(2,-3){2}}
\put(8,4.5){\vector(1,2){2.8}}
\put(8,4.5){\vector(2,-3){1.9}}
\put(10.9,4){\vector(0,-1){2}}
\put(11.1,2){\vector(0,1){2}}
\put(4,1.5){\vector(1,0){2}}
\put(8,1.5){\vector(1,0){2}}
\put(12,13.5){\oval(2,0.5)[r]}
\put(12.5,13.75){\vector(-1,0){0.5}}
\put(12,4.5){\oval(2,0.5)[r]}
\put(12.5,4.75){\vector(-1,0){0.5}}
\put(12,1.5){\oval(2,0.5)[r]}
\put(12.5,1.75){\vector(-1,0){0.5}}
\put(2,10.5){\oval(2,0.5)[l]}
\put(1.5,10.75){\vector(1,0){0.5}}
\put(2,4.5){\oval(2,0.5)[l]}
\put(1.5,4.75){\vector(1,0){0.5}}
\put(2.8,13.4){$x_1$}
\put(6.8,13.4){$x_{10}$}
\put(10.8,13.4){$x_{11}$}
\put(2.8,10.4){$x_2$}
\put(6.8,10.4){$x_{3}$}
\put(10.8,10.4){$x_{12}$}
\put(2.8,7.4){$x_4$}
\put(6.8,7.4){$x_{7}$}
\put(2.8,4.4){$x_5$}
\put(6.8,4.4){$x_{8}$}
\put(10.8,4.4){$x_{13}$}
\put(2.8,1.4){$x_6$}
\put(6.8,1.4){$x_{9}$}
\put(10.8,1.4){$x_{14}$}
\put(1.5,14.2){${\color{red} O_1}$}
\put(5.5,14.2){${\color{red} O_3}$}
\put(9.5,14.2){${\color{red} O_3}$}
\put(1.5,11.2){${\color{red} O_1}$}
\put(5.5,11.2){${\color{red} O_1}$}
\put(9.5,11.2){${\color{red} O_4}$}
\put(1.5,8.2){${\color{red} O_2}$}
\put(5.5,8.2){${\color{red} O_2}$}
\put(1.5,5.2){${\color{red} O_2}$}
\put(5.5,5.2){${\color{red} O_2}$}
\put(9.5,5.2){${\color{red} O_5}$}
\put(1.5,2.2){${\color{red} O_2}$}
\put(5.5,2.2){${\color{red} O_2}$}
\put(9.5,2.2){${\color{red} O_5}$}
\put(4.8,13.7){$u_1$}
\put(8.8,13.8){$u_1$}
\put(8.8,13){$u_1$}
\put(12.8,14){$u_1$}
\put(12.8,5){$u_1$}%
\put(9,12){$u_2$}
\put(10.1,12){$u_2$}
\put(11.2,12){$u_2$}
\put(1,11){$u_1$}
\put(4.8,10.7){$u_2$}
\put(8.8,10.7){$u_1$}
\put(4.8,9.2){$u_1$}
%
\put(9.2,9){$u_2$}
\put(8.8,7.5){$u_2$}
\put(9.3,6){$u_1$}
\put(1,5){$u_2$}
\put(5,3){$u_2$}
\put(8.9,3.3){$u_1$}
\put(10.1,3){$u_2$}
\put(11.2,3){$u_1$}
\put(4.5,1.7){$u_1$}
\put(8.5,1.7){$u_1$}
\put(13,2){$u_2$}
\end{picture}
\caption{Transition System $T_2$ in Example \ref{e3.3} (ii) \label{Fig.fm.5.2}}
\end{figure}

\vskip 5mm

The ASSR of $T_2$ can be obtained by observing Figure \ref{Fig.fm.5.2} as
\begin{align}\label{3.2}
\begin{cases}
x(t+1)=L\sigma(t)x(t),\\
o(t)=Hx(t),
\end{cases}
\end{align}
where,
$$
\begin{array}{ccl}
L&=\d_{14}[&10,2,12,12,0,9,13,14,14,11,10/11,0,13,\\
~&~&13,0,3,0,0,5/9,0,12,12,0,12,12,11,14,14],
\end{array}
$$
$$
H=\d_{5}[1,1,1,2,2,2,2,2,2,3,3,4,5,5].
$$

Under observability equivalence the quotient system can be obtained as
$$
\begin{array}{l}
\con(X_1)=\{x_1,x_2,x_3\},\\
\con(X_2)=\{x_4,x_5,x_6,x_7,x_8,x_9\},\\
\con(X_3)=\{x_{10},x_{11}\},\\
\con(X_4)=\{x_{12}\},\\
\con(X_5)=\{x_{13},x_{14}\}.\\
\end{array}
$$
Figure \ref{Fig.fm.5.3}) shows the quotient system $T_2/\sim$.

\vskip 5mm

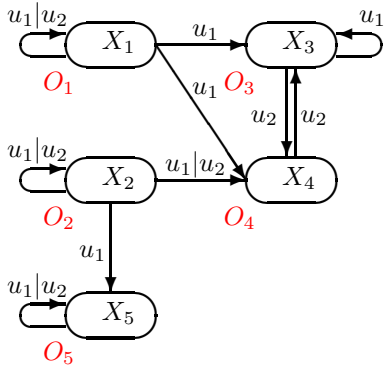
\begin{figure}
\centering
\setlength{\unitlength}{0.6cm}
\begin{picture}(10,9)\thicklines
\put(3,1.5){\oval(2,1)}
\put(3,4.5){\oval(2,1)}
\put(3,7.5){\oval(2,1)}
\put(7,4.5){\oval(2,1)}
\put(7,7.5){\oval(2,1)}
\put(2.8,7.4){$X_1$}
\put(6.8,7.4){$X_3$}
\put(2.8,4.4){$X_2$}
\put(6.8,4.4){$X_4$}
\put(2.8,1.4){$X_5$}
\put(4,7.5){\vector(1,0){2}}
\put(4,4.5){\vector(1,0){2}}
\put(4,7.5){\vector(2,-3){2}}
\put(6.9,7){\vector(0,-1){2}}
\put(7.1,5){\vector(0,1){2}}
\put(3,4){\vector(0,-1){2}}
\put(8,7.5){\oval(2,0.5)[r]}
\put(8.5,7.75){\vector(-1,0){0.5}}
\put(2,7.5){\oval(2,0.5)[l]}
\put(1.5,7.75){\vector(1,0){0.5}}
\put(2,4.5){\oval(2,0.5)[l]}
\put(1.5,4.75){\vector(1,0){0.5}}
\put(2,1.5){\oval(2,0.5)[l]}
\put(1.5,1.75){\vector(1,0){0.5}}
\put(1.5,6.5){${\color{red} O_1}$}
\put(5.5,6.5){${\color{red} O_3}$}
\put(1.5,3.5){${\color{red} O_2}$}
\put(5.5,3.5){${\color{red} O_4}$}
\put(1.5,0.5){${\color{red} O_5}$}
%
\put(4.8,7.7){$u_1$}
\put(4.8,6.4){$u_1$}
\put(4.2,4.7){$u_1|u_2$}
\put(6.1,5.8){$u_2$}
\put(7.2,5.8){$u_2$}
\put(2.3,2.8){$u_1$}
\put(0.7,8){$u_1|u_2$}
\put(0.7,5){$u_1|u_2$}
\put(0.7,2){$u_1|u_2$}
\put(8.5,8){$u_1$}
\end{picture}
\caption{Quotient System $T_2/\sim$ \label{Fig.fm.5.3}}
\end{figure}

\vskip 5mm

Then the ASSR of   $T/\sim$ can be obtained as
\begin{align}\label{3.3}
\begin{cases}
X(t+1)=L_q\sigma(t)X(t),\\
o(t)=H_qX(t),
\end{cases}
\end{align}
where,
$$
L_q=\begin{bmatrix}
1&0&0&0&0&1&0&0&0&0\\
0&1&0&0&0&0&1&0&0&0\\
1&0&1&0&0&0&0&0&1&0\\
1&1&0&0&0&0&1&1&0&0\\
0&1&0&0&1&0&0&0&0&1\\
\end{bmatrix}
$$
$$
H_q=I_5.
$$
\end{itemize}
\end{exa}

The ASSR representation of transition systems is  very convenient in analysis and control design of transition systems. In the following we show how to get the ASSR of quotient systems.

\begin{thm}\label{t3.4}
Consider a transition system
\begin{align}\label{3.4}
\begin{cases}
x(t+1)=Lu(t)x(t),\\
y(t)=Hx(t),
\end{cases}
\end{align}
where $x(t)\in {\cal B}^n$, $y(t)={\cal B}^p$ are Boolean vectors, $u(t)\in \D_m$ is the logical vector,
$L\in {\cal B}_{n\times mn}$ is a Boolean matrix, $H\in {\cal L}_{p\times n}$ is a logical matrix.

Then the quotient system is
\begin{align}\label{3.5}
\begin{cases}
X(t+1)=L_qu(t)X(t),\\
y(t)=H_qX(t),
\end{cases}
\end{align}
where $X_i\in X/\sim$ is the equivalence class of $y_i$, $i\in [1,q]$,
\begin{align}\label{3.6}
L_q=H\times_{\cal B} L\times_{\cal B} (I_m\otimes H^T),
\end{align}
where $\times_{{\cal B}}$ is the Boolean product of matrices;
\begin{align}\label{3.7}
H_q=I_p.
\end{align}
\end{thm}

\noindent {\it Proof.}  Since $X_i$ is the equivalence class of $y_i$, that is, $x\in \con(X_i)\Leftrightarrow y(x)=O_i$, then formula (\ref{fm.5.7}) is obvious.

To prove (\ref{3.6}), denote
$$
L_q=[L_q^1,L_q^2,\cdots,L_q^m],
$$
where $L_q^j=L_q\d_m^j$, $j\in[1,m]$. Similarly,  $L$ is divided into
$$
L=[L^1,L^2,\cdots,L^m].
$$

First, note that according to (\ref{3.5}),  $L_q^j$ is the transition matrix of subsets
$(\con(X_1),\con(X_2),\cdots,\con(X_p))$ to subsets $(\con(X_1),\con(X_2),\cdots,\con(X_p))$ when the control $u=\d_m^j$.

Second, according to the structure it is clear that  $\Row_j(H)$ is the index function of $\con(X_j)$, $j\in [1,p]$.

Recall the transition matrix of one set of subsets to another set of subsets (refer to Chapter 4 of Volume 2, or \cite{che18}), it is clear that
when $u=\d_m^j$ the transition matrix of
set of subsets $(\con(X_1),\con(X_2),\cdots,\con(X_p))$ to set of subsets $(\con(X_1),\con(X_2),\cdots,\con(X_p))$ is
$$
H\times_{{\cal B}} L^j \times_{{\cal B}} H^T.
$$
Hence we have
\begin{align}\label{3.7a}
L_q^j=H\times_{{\B}} L^j \times_{{\B}} H^T,\quad j\in [1,m].
\end{align}
Put $m$ equations of (\ref{3.7a}) together yields (\ref{3.6}).
\hfill $\Box$

A transition system without control $U$ is called an autonomous transition system.

\begin{cor}\label{c3.5} Assume $T=(X,\Sigma,O,h)$ is an autonomous transition system with ASSR
\begin{align}\label{3.8}
\begin{cases}
x(t+1)=Mx(t),\\
y(t)=Hx(t).
\end{cases}
\end{align}
Then the quotient system under output equivalence relation is
\begin{align}\label{3.9}
\begin{cases}
X(t+1)=M_qX(t),\\
y(t)=H_qX(t),
\end{cases}
\end{align}
{where}
\begin{align}\label{3.10}
\begin{array}{l}
M_q=H\times_{\B} M \times_{\B} H^T,\\
H_q=I_p.
\end{array}
\end{align}
\end{cor}

Denote the set of output trajectories of $T$ by ${\bf L}_T(x)$. (If $T$ is not determinant the trajectories are not unique.) If $S_0\subset X$, then
$$
{\bf L}_T(S_0)=\bigcup_{x\in S_0} {\bf L}_T(x).
$$

Let  $T$ be a transition system and $T/\sim$ is its quotient system. Then according to the definition it is clear that
\begin{align}\label{3.11}
{\bf L}_T(\con(X_i))\subset {\bf L}_{T/\sim}(X_i).
\end{align}

That is, each output sequence can be produced by its quotient system. Because of this, the quotient system $T/\sim$ is said to be a simulation of the original $T$.

The following simple example shows an output sequence of $T/\sim$ may not be obtained from the original $T$.

\begin{exa}\label{e3.6}

Observe a $T$ depicted in Figure \ref{Fig.fm.5.4}.

 \vskip 5mm

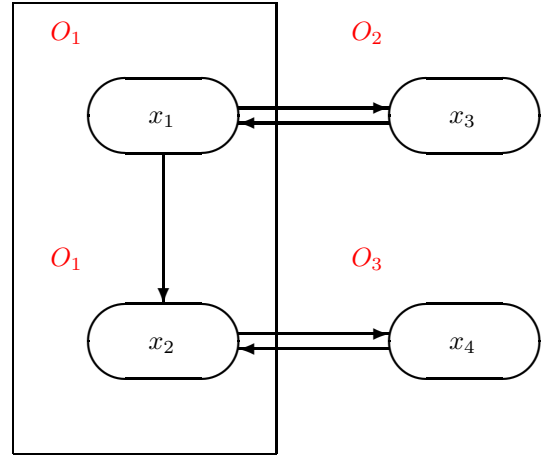
\begin{figure}
\centering
\setlength{\unitlength}{1cm}
\begin{picture}(9,6)\thicklines
\put(3,1.5){\oval(2,1)}
\put(3,4.5){\oval(2,1)}
\put(7,1.5){\oval(2,1)}
\put(7,4.5){\oval(2,1)}
\put(2.8,1.4){$x_2$}
\put(6.8,1.4){$x_4$}
\put(2.8,4.4){$x_1$}
\put(6.8,4.4){$x_3$}
\put(4,1.6){\vector(1,0){2}}
\put(6,1.4){\vector(-1,0){2}}
\put(4,4.6){\vector(1,0){2}}
\put(6,4.4){\vector(-1,0){2}}
\put(3,4){\vector(0,-1){2}}
\put(1.5,5.5){${\color{red} O_1}$}
\put(5.5,5.5){${\color{red} O_2}$}
\put(1.5,2.5){${\color{red} O_1}$}
\put(5.5,2.5){${\color{red} O_3}$}
\thinlines
\put(1,0){\line(1,0){3.5}}
\put(1,0){\line(0,1){6}}
\put(4.5,6){\line(-1,0){3.5}}
\put(4.5,6){\line(0,-1){6}}
\end{picture}
\caption{Transition system of Example \ref{e3.6} \label{Fig.fm.5.4}}
\end{figure}

\vskip 5mm

It is clear that  $\L_{T/\sim}(X_3)= O_3O_1O_2\cdots$ is an output trajectory, which is not an output trajectory of $\L_{T}(\con(X_3))$.

\end{exa}

\begin{dfn}\label{d3.7} \cite{bel17} Consider a transition system $T=(X,U,\Sigma,O,h)$. Assume $x_1\sim x_2$ are (output) equivalence, if for each control $u\in U$ with $x_1'\in \Sigma(x_1,u)$ there exists an $x_2'\in \Sigma(x_2,u)$ such that $x_1'\sim x_2'$, then $x_1\approx x_2$. If for any $x_1\sim x_2$, we have $x_1\approx x_2$, then  $T/\sim$ is called a bi-simulation of $T$, denoted by $T/\approx$.
\end{dfn}

According to Definition \ref{d3.7}, bi-simulation means for all $X_i\in T/\sim$ we have
\begin{align}\label{3.12}
{\bf L}_T(\con(X_i)) = {\bf L}_{T/\approx}(X_i).
\end{align}

The following is an important property of bi-simulation:

\begin{prp}\label{p3.8} \cite{bel17} Consider a transition system $T=(X,U,\Sigma,O,h)$.
\begin{itemize}
\item[(i)] If $T/\sim$ is a deterministic system, then $T/\sim=T/\approx$ is a bi-simulation.
\item[(ii)] If $T$ is a deterministic system, that $T/\sim=T/\approx$ is a bi-simulation, if and only if, $T/\sim$ is deterministic.
\end{itemize}
\end{prp}

\section{Aggregation via Bi-simulation}

\begin{dfn}\label{d4.1}
\begin{itemize}
\item[(i)] A $k$-valued (autonomous) networked system can be described as
\begin{align}\label{4.1}
\begin{cases}
x_1(t+1)=f_1(x_1(t),\cdots,x_n(t)),\\
x_2(t+1)=f_2(x_1(t),\cdots,x_n(t)),\\
\vdots\\
x_n(t+1)=f_n(x_1(t),\cdots,x_n(t)),\\
\end{cases}
\end{align}
where $x_i(t)\in {\cal D}_k$, $i\in [1,n]$ are states,  $f_i:{\cal D}_k^n\ra {\cal D}_k$, $i\in [1,n]$ are state transition functions.

\item[(ii)] A $k$ valued networked control system can be described as
\begin{align}\label{4.2}
\begin{array}{l}
\begin{cases}
x_1(t+1)=f_1(x_1(t),\cdots,x_n(t), u_1(t),\cdots,u_m(t)),\\
x_2(t+1)=f_2(x_1(t),\cdots,x_n(t), u_1(t),\cdots,u_m(t)),\\
\vdots,\\
x_n(t+1)=f_n(x_1(t),\cdots,x_n(t), u_1(t),\cdots,u_m(t)),\\
\end{cases}\\
~~~y_j(t)=h_j(x_1(t),\cdots,x_n(t)),\quad j\in [1,p],
\end{array}
\end{align}
where $x_i(t)\in {\cal D}_k$, $i\in [1,n]$ are states,  $u_s(t)\in {\cal D}_k$, $s\in [1,m]$ are inputs (or controls),
 $y_j(t)\in {\cal D}_k$, $j\in [1,p]$ are outputs (or observations), $f_i:{\cal D}_k^n\times {\cal D}_k^m\ra {\cal D}_k$, $i\in [1,n]$ are state updating functions, $h_j:{\cal D}_k^n\ra {\cal D}_k$, $j\in [1,p]$ are output functions.
\end{itemize}
\end{dfn}

Consider networked system (\ref{4.1}), usually only part of $\{x_i\;|\;i\in [1,n]\}$ appear into $f_i$. That is,
$$
f_i=f_i(x_{r_1},x_{r_2},\cdots,x_{r_{n_i}}),\quad i\in [1,n].
$$
Similarly, for system (\ref{4.2}),
$$
\begin{array}{l}
f_i=f_i(x_{r_1},x_{r_2},\cdots,x_{r_{n_i}},u_{s_1},u_{s_2},\cdots,u_{s_{m_i}}),\quad i\in [1,n],\\
h_j=h_j(x_{\ell_1},x_{\ell_2},\cdots,x_{\ell_{p_i}}),\quad j\in [1,p].
\end{array}
$$

\begin{dfn}\label{d4.2}
\begin{itemize}
\item[(i)] A directed graph $(N,E)$  is called a network graph of (\ref{4.1}), if
$$
\begin{array}{l}
N=\{x_1,x_2,\cdots,x_n\}\\
(x_{\a}, x_{\b})\in E \Leftrightarrow x_{\a}\in f_{\b}.
\end{array}
$$
\item[(ii)] A directed graph $(N,E)$  is called a network graph of (\ref{4.2}), if
$$
\begin{array}{l}
N=\{x_1,x_2,\cdots,x_n,u_1,u_2,\cdots,u_m,o_1,o_2,\cdots,o_p\}\\
\begin{cases}
(x_{\a}, x_{\b})\in E \Leftrightarrow x_{\a}\in f_{\b},\\
(u_{\a}, x_{\b})\in E \Leftrightarrow u_{\a}\in f_{\b},\\
(x_{\a}, o_{\b})\in E \Leftrightarrow x_{\a}\in h_{\b}.\\
\end{cases}
\end{array}
$$
\end{itemize}
\end{dfn}

\begin{exa}\label{e4.3}
Consider a Boolean network, which has its network equation as
\begin{align}\label{4.3}
\begin{array}{l}
\begin{cases}
x_1(t+1)=\neg x_1(t),\\
x_2(t+1)=x_1(t)\wedge x_3(t),\\
x_3(t+1)=x_3(t)\vee x_4(t),\\
x_4(t+1)=x_3(t)\ra x_5(t),\\
x_5(t+1)=x_2(t)\bar{\vee} x_4(t),\\
x_6(t+1)=x_4(t)\lra x_6(t),\\
\end{cases}\\
~~y(t)=x_6(t).
\end{array}
\end{align}
Then its network graph is depicted  by Figure \ref{Fig.4.1}.

\vskip 5mm

\begin{figure}
\centering
\setlength{\unitlength}{1cm}
\begin{picture}(5,8)(-1,-1)\thicklines
\put(0,3){\vector(1,-1){1}}
\put(0,3){\vector(1,1){1}}
\put(2,3){\vector(-1,-1){1}}
\put(1,6){\vector(0,-1){2}}
\put(1,2){\vector(0,-1){2}}
\put(1,0){\vector(1,0){1.8}}
%
\put(1,4){\vector(1,-1){1}}
\put(1,2){\vector(-1,1){1}}
\put(1,2){\vector(1,1){1}}
%
\put(1.2,5.8){$x_1$}
\put(1.2,4.2){$x_2$}
\put(-0.4,2.8){$x_3$}
\put(2.1,2.8){$x_5$}
\put(1.2,1.8){$x_4$}
\put(0.5,0.2){$x_6$}
\put(1.5,0.2){$y=x_6$}
%
\thinlines
\put(-0.5,1){\framebox(3,4)}
\end{picture}
\caption{Boolean network (\ref{4.3}) \label{Fig.4.1}}
\end{figure}
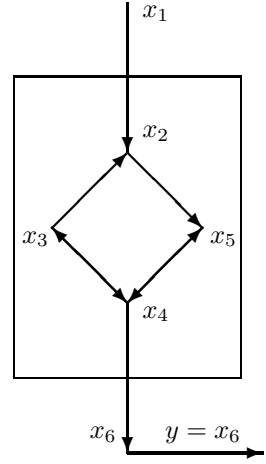

\vskip 5mm
\end{exa}

\begin{dfn}\label{d4.4}
Consider an autonomous network $\Sigma$ with its network graph $(N,E)$, where $N=\{x_1,x_2,\cdots,x_n\}$.
Let $A=\{x_{a_1},\cdots,x_{a_{n_A}}\}\subset N$, where $\{a_1,a_2,\cdots,a_{n_A}\}\subset [1,n]$.
\begin{itemize}
\item[(i)] If $(x_i,x_j)\in E$, $x_i\in A^c$, $x_j\in A$, $x_i$ is called a block input of $A$.
\item[(ii)] If $(x_i,x_j)\in E$, $x_i\in A$, $x_j\in A^c$, $x_i$ is called a formal output of $A$.
\end{itemize}
$A$ is called an {aggregate-able block} if there does not exist $(x_i,x_j)\in E$ such that $x_i$ is a formal input and $x_j$ is a formal output.
\end{dfn}

\begin{prp}\label{p4.5}
Assume $A\subset N$ is an aggregate-able block with
$\{x_{i_1},\cdots,x_{i_{\a}}\}$ as its block inputs, and $\{\{x_{j_1},\cdots,x_{j_{\b}}\}$ as its block outputs.
Then the dynamic subnetwork of $A$ can be expressed as a controlled network $\Sigma_A$ with block control
$$
v_{\ell}:=x_{i_\ell},\quad \ell\in[1,\a],
$$
and block output
$$
y_{\ell}:=x_{j_{\ell}},\quad \ell\in [1,\b].
$$
Replacing block $A$ in $\Sigma$ by this block control system $\Sigma_A$ does not affect the dynamics of the rest part of $\Sigma$.
\end{prp}

\noindent{\it Proof.} From the construction it is clear that this replacement does not change anything for $N\backslash A$ except changing some variable names.
\hfill $\Box$

\begin{exa}\label{e4.6} Recall Example \ref{e4.3}. It is easy to calculate that the ASSR of (\ref{4.3}) is
\begin{align}\label{4.4}
x(t+1)=Mx(t),
\end{align}
where $x(t)=\ltimes_{i=1}^6x_i(t)$,
$$
\begin{array}{ccl}
M&=\d_{64}[&35,36,39,40,34,33,38,37,\\
~&~        &51,52,51,52,58,57,58,57,\\
~&~        &33,34,37,38,36,35,40,39,\\
~&~        &49,50,49,50,60,59,60,59,\\
~&~        &19,20,23,24,18,17,22,21,\\
~&~        &19,20,19,20,26,25,26,25,\\
~&~        &17,18,21,22,20,19,24,23,\\
~&~        &17,18,17,18,28,27,28,27].
\end{array}
$$

Next, we consider $A=\{x_2,x_3,x_4,x_5\}\subset N$. It is easy to verify that $A$ is an aggregate-able set with
block input $v=x_1$ and block output $o=x_4$. Then $\Sigma_A$ can be expressed as
\begin{align}\label{4.5}
\begin{cases}
z(t+1)=L_Av(t)z(t),\\
y(t)=Hz(t),
\end{cases}
\end{align}
where $(z_1(t),z_2(t),z_3(t),z_4(t)\}=\{x_2(t),x_3(t),x_4(t),x_5(t)\}$, $z(t)=\ltimes_{i=2}^5x_i(t)=\ltimes_{i=1}^4z_i(t)$, $v(t)=x_1(t)$, $y(t)=z_3(t)=x_4(t)$,
$$
\begin{array}{ccl}
L_A&=\d_{16}[& 2, 4, 1, 3,10,10,13,13,\\
~&~        & 1, 3, 2, 4, 9, 9,14,14,\\
~&~        &10,12, 9,11,10,10,13,13,\\
~&~        & 9,11,10,12, 9, 9,14,14].
\end{array}
$$
$$
H=\d_2[1,1,2,2,1,1,2,2,1,1,2,2,1,1,2,2].
$$
\end{exa}

\begin{dfn}\label{d4.7} Consider a networked system $\Sigma$ with its network graph $(N,E)$. Assume $A\subset N$ is an aggregate-able subset, and the aggregated (block control) system is $\Sigma_A$. $A$ is said to be aggregated by its simulation, if $\Sigma_A$ is replaced by its quotient system $\Sigma_A/\sim$.
\end{dfn}

\begin{exa}\label{e4.8} Recall Example \ref{e4.3} (or, Example \ref{e4.6}). According to Theorem \ref{t3.4} and
Using the results in Example (\ref{e4.6}), the $\Sigma_A/\sim$ is
\begin{align}\label{4.6}
\begin{cases}
w(t+1)=Lv(t)w(t),\\
y(t)=w(t),
\end{cases}
\end{align}
where
$v(t)=x_1(t)$, $w(t)=y(t)=x_4(t)$,
$$
\begin{array}{ccl}
L&=&H\times_{{\cal B}}L_A\times_{{\cal B}}(I_2\otimes H^T)\\
~&=&\begin{bmatrix}1&1&1&1\\1&1&1&1\end{bmatrix} 
\end{array}
$$
\end{exa}

\begin{rem}\label{r4.9}
\begin{itemize}
\item[(i)] From Example \ref{e4.8} one sees that when a block of a networked system is aggregated by simulation its size is reduced. Unfortunately,  the resulting overall system is only a transition system, which may not be determinant.
\item[(ii)] The aggregation by simulation proposed for networked system is also applicable to networked control system with an obvious extension. Precisely, it can be done for each original control $u_s=\d_{m}^s$ separately.
\item[(iii)] For a simulation aggregation of $A\subset N$, if $\Sigma_A/\sim=\Sigma_A/\approx$, the aggregation is called a bi-simulation.
\end{itemize}
\end{rem}

\begin{prp}\label{p4.10}  Assume $\Sigma$ is a networked system $\sigma$ with its network graph $(N,E)$, $A\subset N$ is an aggregate-able subset,  $\Sigma_A/\sim=\Sigma_A/\approx$  is  a bi-simulation, then the aggregation does not affect the dynamics of the overall system.
\end{prp}

\noindent{\it Proof.} When $\Sigma_A/\sim$ is a bisimulation of $\Sigma_A$, according to Proposition \ref{p3.8}, $\Sigma_A/\sim$ is deterministic; therefore, the values of input nodes and output nodes characterize the states of $\Sigma_A$ at each time and uniquely determines the input and output at next moment, hence the inner structure of the subnetwork can be neglected. 
\hfill $\Box$

\begin{exa}\label{e4.11}

Consider a Boolean control network $\Sigma$ whose network graph is depicted  by Figure \ref{Fig.4.2}.

\vskip 5mm
\begin{figure}
\centering
\setlength{\unitlength}{0.8cm}
\begin{picture}(8,6)(0,-0.5)\thicklines
\put(2,0){\framebox(4.5,5.5)}
\put(2.5,0.5){\framebox(3,1.5){$\Sigma_A$}}
\put(2.5,3){\framebox(3,1.5){$\Sigma_{N\backslash A}$}}
\put(2.2,5){$\Sigma$}
\put(0,2.5){\vector(1,0){2}}
\put(1,2.5){\line(0,-1){1.25}}
\put(1,1.25){\vector(1,0){1.5}}
\put(6.5,2.5){\vector(1,0){2}}
\put(5.5,1.25){\line(1,0){2}}
\put(7.5,1.25){\vector(0,1){1.25}}
\put(3.2,3){\vector(0,-1){1}}
\put(4.8,2){\vector(0,1){1}}
\put(0.5,2.7){$u$}
\put(7,2.7){$y$}
\put(1.2,1.5){$u_A$}
\put(7.6,1.7){$y_A$}
\put(3.3,2.5){$x_i(t)$}
\put(4.9,2.5){$x_{i+\mu}(t)$}
\end{picture}
\caption{A Boolean control network $\Sigma$ \label{Fig.4.2}}
\end{figure}
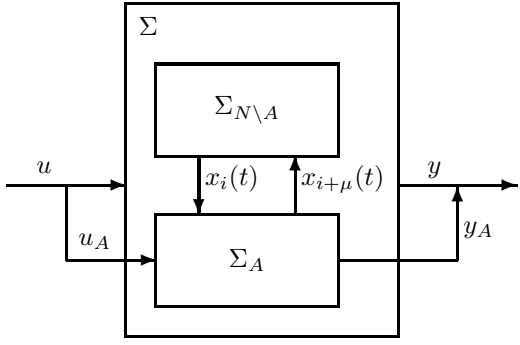

\vskip 5mm

Assume the set of nodes of $\Sigma$ is $N$ and $A\subset N$, where
$$
A=\{x_{i+1},x_{i+1},\cdots,x_{i+\mu}\},\quad \mu>1.
$$

The dynamic equations of $A$, denoted by $\Sigma_A$,  are
\begin{align}\label{4.7}
\begin{array}{l}
\begin{cases}
x_{i+1}(t+1)&=\left[(x_i(t)\bar{\vee} x_{i+1}(t))\wedge u(t)\right]\\
~&\vee \left[(x_i(t)\lra x_{i+1}(t))\wedge \neg u(t)\right],\\
x_{i+2}(t+1)&=\left[(x_{i+1}(t)\bar{\vee} x_{i+2}(t))\wedge u(t)\right]\\
~&\vee \left[(x_{i+1}(t)\lra x_{i+2}(t))\wedge \neg u(t)\right],\\
\vdots\\
x_{i+\mu}(t+1)&=\left[(x_{i+\mu-1}(t)\bar{\vee} x_{i+\mu}(t))\wedge u(t)\right]\\
~&\vee \left[(x_{i+\mu-1}(t)\lra x_{i+\mu}(t))\wedge \neg u(t)\right].\\
\end{cases}\\
y(t)=x_{i+\mu}(t).
\end{array}
\end{align}
Using Theorem \ref{t3.4}, the quotient system  $\Sigma_A/\sim$ is calculated as
\begin{align}\label{4.8}
y(t+1)=\d_2[2,1,1,2,1,2,2,1]u(t)v(t)y(t),
\end{align}
where $y(t)=x_{i+\mu}(t)$, $v(t)=x_i(t)$.  According to Proposition \ref{p3.8}, $\Sigma_A/\sim=\Sigma_A/\approx$ is a bi-simulation. According to

Proposition \ref{p4.10},  (\ref{4.7}) can be replaced by (\ref{4.8}), which does not affect the input-output mapping of the overall system $\Sigma$.

\end{exa}

\section{Aggregated Simulation}

\begin{figure}
\centering
\setlength{\unitlength}{0.8cm}
\begin{picture}(7,3)\thicklines
\put(2,0.5){\framebox(3,2){$\Sigma_A$}}
\put(0.5,1){\vector(1,0){2}}
\put(0.5,2){\vector(1,0){2}}
\put(4.5,1){\vector(1,0){2}}
\put(4.5,2){\vector(1,0){2}}
\put(0.7,1.2){$v_{\a}$}
\put(0.7,2.2){$v_1$}
\put(1.5,1.3){$\vdots$}
\put(5.7,1.2){$y_{\b}$}
\put(5.7,2.2){$y_1$}
\put(5.5,1.3){$\vdots$}
\end{picture}
\caption{An Aggregated Block \label{Fig.5.1}}
\end{figure}
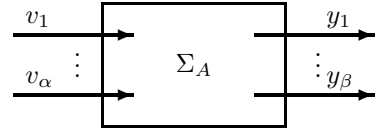

Consider a $k$-valued network $\Sigma$ with its network graph $(N,E)$. Assume a set of sub-nodes $A\subset N$ is an aggregate-able and output decoupled block depicted in Fig. \ref{Fig.5.1}.

\vskip 5mm

\vskip 5mm

Moreover, assume the dynamic equations of $A$ with block inputs $v_1,\cdots,v_{\a}\in N\backslash{A}$ and block outputs $y_1,\cdots,y_{\b}\in A$
is
\begin{align}\label{5.1}
\begin{cases}
z(t+1)=L_Au(t)v(t)z(t),\\
y(t)=H_Az(t).
\end{cases}
\end{align}
Then its quotient system under output equivalence, denoted by $\Sigma_A/\sim$, is
\begin{align}\label{5.2}
y(t+1)=\tilde{L}_Au(t)v(t)y(t),
\end{align}
where $\tilde{L}_A=H_A\times_{{\cal B}} L_A\times (I_{k^{m+\a}}\otimes H^T_A)\in {\cal B}_{k^{\b}\times k^{m+\a+\b}}:={\cal B}_{\xi\times \eta}$, where
$$
\xi:=k^{\b},\quad \eta:=k^{m+\a+\b}.
$$

If $\tilde{L}_A$ is a logical matrix, which means $\Sigma_A/\sim =\Sigma_A/\approx$ is deterministic, then the bisimulation is applicable. That is, the block $A$ can be replaced by $\Sigma_A/\sim$, which does not affect the input-output relationship of the overall network $\Sigma$. In the following assume $\Sigma_A$ is not deterministic.

In the following we propose a method called the aggregation by simulation, which is an approximation by probabilistic networks.

\begin{dfn}\label{d5.1} Given a $k$-valued network $\Sigma$ with network graph $(N,E)$. A subset $A\subset N$ is an aggregate-able and output decoupled block, which is described as above. The aggregation by simulation is defined as follows:
\begin{itemize}
\item Step 1. Construct the quotient system $\Sigma_A/\sim$.
\item Step 2. Using $\Sigma_A/\sim$ to build a probabilistic network $\Sigma_A^P$ as follows: Set
$$
M_A:=H_AL_A(I_{k^{m+\a}}\otimes H^T_A)=(m_{i,j})\in {\cal M}_{\xi\times \eta}.
$$
Denote
$$
m_j:=\dsum_{i=1}^{\xi}m_{i,j},\quad j\in [1,\eta].
$$
Define a probabilistic system, denoted by $\Sigma_A^P$ as follows:
$$
y(t+1)=M^{i_1,i_2,\cdots,i_{\eta}}u(t)v(t)y(t),\quad i_j\in [1,\xi],\;j\in [1,\eta],
$$
where
$$
M^{i_1,i_2,\cdots,i_{\eta}}=\d_{\xi}[i_1,i_2,\cdots,i_{\eta}],
$$
with probability
$$
p_{i_1,i_2,\cdots,i_{\eta}}=\frac{\prod_{j=1}^\eta m_{i_j,j}}{\prod_{j=1}^\eta m_j}.
$$
\item Step 3. Replace $\Sigma_A$ by $\Sigma_A^P$.
\end{itemize}
\end{dfn}

\begin{exa}\label{e5.2} Recall Example \ref{e4.3} (or, Examples \ref{e4.6} and \ref{e4.8}). Using  Definition \ref{d5.1}, the system $\Sigma_A^P/\sim$ can be calculated as follows:

\begin{align}\label{5.3}
\begin{array}{ccl}
M_A&=&HL_A(I_2\otimes H^T)\\
~&=&\begin{bmatrix}6&6&6&6\\2&2&2&2\end{bmatrix}.
\end{array}
\end{align}

Then the simulation-aggregation is using the following probabilistic network to replace $A$:
\begin{align}\label{5.4}
z(t+1)=L_A^Pv(t)z(t),
\end{align}
where
$$
L_A^P=\begin{bmatrix}
2/3&2/3&2/3&2/3\\
1/3&1/3&1/3&1/3
\end{bmatrix}.
$$
\end{exa}

\section{Aggregated Simulation of a Biological System}

This section considers an example of T-cell receptor kinetics which was originally modeled in \cite{kla06}. The following model is copied from \cite{zhu22} and ignoring nodes' physical meanings.

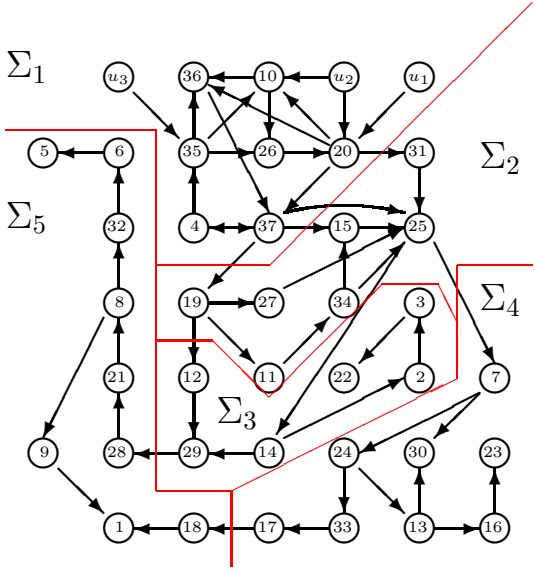
\begin{figure}
\centering
\setlength{\unitlength}{1cm}
\begin{picture}(7,7)(-0.5,-0.5)\thicklines
\put(1,0){\circle{0.4}}
\put(0.95,-0.05){\begin{tiny}$1$\end{tiny}}
\put(2,0){\circle{0.4}}
\put(1.85,-0.05){\begin{tiny}$18$\end{tiny}}
\put(3,0){\circle{0.4}}
\put(2.85,-0.05){\begin{tiny}$17$\end{tiny}}
\put(4,0){\circle{0.4}}
\put(3.85,-0.05){\begin{tiny}$33$\end{tiny}}
\put(5,0){\circle{0.4}}
\put(4.85,-0.05){\begin{tiny}$13$\end{tiny}}
\put(6,0){\circle{0.4}}
\put(5.85,-0.05){\begin{tiny}$16$\end{tiny}}
\put(1.8,0){\vector(-1,0){0.6}}
\put(2.8,0){\vector(-1,0){0.6}}
\put(3.8,0){\vector(-1,0){0.6}}
\put(5.2,0){\vector(1,0){0.6}}
\put(0.2,0.8){\vector(1,-1){0.6}}
\put(4,0.8){\vector(0,-1){0.6}}
\put(4.2,0.8){\vector(1,-1){0.6}}
\put(5,0.2){\vector(0,1){0.6}}
\put(6,0.2){\vector(0,1){0.6}}
\put(0,1){\circle{0.4}}
\put(-0.05,0.95){\begin{tiny}$9$\end{tiny}}
\put(1,1){\circle{0.4}}
\put(0.85,0.95){\begin{tiny}$28$\end{tiny}}
\put(2,1){\circle{0.4}}
\put(1.85,0.95){\begin{tiny}$29$\end{tiny}}
\put(3,1){\circle{0.4}}
\put(2.85,0.95){\begin{tiny}$14$\end{tiny}}
\put(4,1){\circle{0.4}}
\put(3.85,0.95){\begin{tiny}$24$\end{tiny}}
\put(5,1){\circle{0.4}}
\put(4.85,0.95){\begin{tiny}$30$\end{tiny}}
\put(6,1){\circle{0.4}}
\put(5.85,0.95){\begin{tiny}$23$\end{tiny}}
\put(0.8,2.8){\vector(-1,-2){0.8}}
\put(1,1.2){\vector(0,1){0.6}}
\put(1.8,1){\vector(-1,0){0.6}}
\put(2,1.8){\vector(0,-1){0.6}}
\put(2.8,1){\vector(-1,0){0.6}}
\put(4.8,3.8){\vector(-2,-3){1.7}}
\put(3.2,1.2){\vector(2,1){1.6}}
\put(5.8,1.8){\vector(-2,-1){1.6}}
\put(5.8,1.8){\vector(-1,-1){0.6}}
\put(1,2){\circle{0.4}}
\put(0.85,1.95){\begin{tiny}$21$\end{tiny}}
\put(2,2){\circle{0.4}}
\put(1.85,1.95){\begin{tiny}$12$\end{tiny}}
\put(3,2){\circle{0.4}}
\put(2.85,1.95){\begin{tiny}$11$\end{tiny}}
\put(4,2){\circle{0.4}}
\put(3.85,1.95){\begin{tiny}$22$\end{tiny}}
\put(5,2){\circle{0.4}}
\put(4.95,1.95){\begin{tiny}$2$\end{tiny}}
\put(6,2){\circle{0.4}}
\put(5.95,1.95){\begin{tiny}$7$\end{tiny}}
\put(1,2.2){\vector(0,1){0.6}}
\put(2,2.8){\vector(0,-1){0.6}}
\put(2.2,2.8){\vector(1,-1){0.6}}
\put(3.2,2.2){\vector(1,1){0.6}}
\put(4.8,2.8){\vector(-1,-1){0.6}}
\put(5.2,3.8){\vector(1,-2){0.8}}
\put(5,2.2){\vector(0,1){0.6}}
\put(1,3){\circle{0.4}}
\put(0.95,2.95){\begin{tiny}$8$\end{tiny}}
\put(2,3){\circle{0.4}}
\put(1.85,2.95){\begin{tiny}$19$\end{tiny}}
\put(3,3){\circle{0.4}}
\put(2.85,2.95){\begin{tiny}$27$\end{tiny}}
\put(4,3){\circle{0.4}}
\put(3.85,2.95){\begin{tiny}$34$\end{tiny}}
\put(5,3){\circle{0.4}}
\put(4.95,2.95){\begin{tiny}$3$\end{tiny}}
\put(1,3.2){\vector(0,1){0.6}}
\put(2.8,3.8){\vector(-1,-1){0.6}}
\put(2.2,3){\vector(1,0){0.6}}
\put(3.2,3.2){\vector(2,1){1.6}}
\put(4,3.2){\vector(0,1){0.6}}
\put(4.2,3.2){\vector(1,1){0.6}}
\put(1,4){\circle{0.4}}
\put(0.85,3.95){\begin{tiny}$32$\end{tiny}}
\put(2,4){\circle{0.4}}
\put(1.95,3.95){\begin{tiny}$4$\end{tiny}}
\put(3,4){\circle{0.4}}
\put(2.85,3.95){\begin{tiny}$37$\end{tiny}}
\put(4,4){\circle{0.4}}
\put(3.85,3.95){\begin{tiny}$15$\end{tiny}}
\put(5,4){\circle{0.4}}
\put(4.85,3.95){\begin{tiny}$25$\end{tiny}}
\put(1,4.2){\vector(0,1){0.6}}
\put(2,4.2){\vector(0,1){0.6}}
\put(2.2,4){\vector(1,0){0.6}}
\put(2.8,4){\vector(-1,0){0.6}}
\put(3.2,4){\vector(1,0){0.6}}
\put(4.2,4){\vector(1,0){0.6}}
\qbezier(3.2,4.2)(4,4.4)(4.8,4.2)
\put(4.6,4.25){\vector(4,-1){0.2}}
\put(2.2,5.8){\vector(1,-2){0.8}}
\put(3.8,4.8){\vector(-1,-1){0.6}}
\put(5,4.8){\vector(0,-1){0.6}}
\put(0,5){\circle{0.4}}
\put(-0.05,4.95){\begin{tiny}$5$\end{tiny}}
\put(1,5){\circle{0.4}}
\put(0.95,4.95){\begin{tiny}$6$\end{tiny}}
\put(2,5){\circle{0.4}}
\put(1.85,4.95){\begin{tiny}$35$\end{tiny}}
\put(3,5){\circle{0.4}}
\put(2.85,4.95){\begin{tiny}$26$\end{tiny}}
\put(4,5){\circle{0.4}}
\put(3.85,4.95){\begin{tiny}$20$\end{tiny}}
\put(5,5){\circle{0.4}}
\put(4.85,4.95){\begin{tiny}$31$\end{tiny}}
\put(0.8,5){\vector(-1,0){0.6}}
\put(1.2,5.8){\vector(1,-1){0.6}}
\put(2.2,5){\vector(1,0){0.6}}
\put(2,5.2){\vector(0,1){0.6}}
\put(2.2,5.2){\vector(1,1){0.6}}
\put(3.8,5.1){\vector(-2,1){1.6}}
\put(3,5.8){\vector(0,-1){0.6}}
\put(3.2,5){\vector(1,0){0.6}}
\put(4.2,5){\vector(1,0){0.6}}
\put(3.8,5.2){\vector(-1,1){0.6}}
\put(4.8,5.8){\vector(-1,-1){0.6}}
\put(4,5.8){\vector(0,-1){0.6}}
\put(1,6){\circle{0.4}}
\put(0.85,5.95){\begin{tiny}$u_3$\end{tiny}}
\put(2,6){\circle{0.4}}
\put(1.85,5.95){\begin{tiny}$36$\end{tiny}}
\put(3,6){\circle{0.4}}
\put(2.85,5.95){\begin{tiny}$10$\end{tiny}}
\put(4,6){\circle{0.4}}
\put(3.85,5.95){\begin{tiny}$u_2$\end{tiny}}
\put(5,6){\circle{0.4}}
\put(4.85,5.95){\begin{tiny}$u_1$\end{tiny}}
\put(2.8,6){\vector(-1,0){0.6}}
\put(3.8,6){\vector(-1,0){0.6}}
\put(-0.5,6){{\large $\Sigma_1$}}
\put(-0.5,4){{\large $\Sigma_5$}}
\put(5.8,4.8){{\large $\Sigma_2$}}
\put(2.3,1.4){{\large $\Sigma_3$}}
\put(5.8,2.9){{\large $\Sigma_4$}}
\thinlines
{\color{red}
\put(-0.5,5.3){\line(1,0){2}}
\put(1.5,5.3){\line(0,-1){4.8}}
\put(1.5,0.5){\line(1,0){1}}
\put(2.5,0.5){\line(2,1){3}}
\put(5.5,2){\line(0,1){1.5}}
\put(5.5,3.5){\line(1,0){1}}
\put(1.5,3.5){\line(1,0){1.5}}
\put(3,3.5){\line(1,1){3.5}}
\put(1.5,2.5){\line(1,0){0.75}}
\put(2.25,2.5){\line(1,-1){0.75}}
\put(3,1.75){\line(1,1){1.5}}
\put(4.5,3.25){\line(1,0){0.75}}
\put(5.25,3.25){\line(1,-2){0.25}}
\put(2.5,0.5){\line(0,-1){1}}
}
\end{picture}
\caption{T-cell receptor kinetics \label{Fig.6.1}}
\end{figure}

The dynamics of network of T-cell receptor kinetics with $37$ nodes and $3$ controls, depicted in Fig. \ref{Fig.6.1}, is described in (\ref{6.1}).

\vskip 5mm

\vskip 5mm

\begin{align}\label{6.1}
\begin{array}{ll}
x_1(*)=x_9\wedge x_{18},&x_2(*)=x_{14},\\
x_3(*)=x_2,&x_4(*)=x_{37},\\
x_5(*)=x_6,&x_6(*)=x_{32},\\
x_7(*)=x_{25},&x_8(*)=x_{21},\\
x_9(*)=x_8,&x_{10}(*)=(x_{20}\wedge u_2)\\
x_{11}(*)=x_{19},& ~~\vee(x_{35}\wedge u_2),\\
x_{12}(*)=x_{19},& x_{13}(*)=x_{24},\\
x_{14}(*)=x_{25},& x_{15}(*)=x_{34}\wedge x_{37},\\
x_{16}(*)=\overline{x_{13}},& x_{17}(*)=x_{33},\\
x_{18}(*)=x_{17},& x_{19}(*)=x_{37},\\
x_{20}(*)=\overline{x_{26}}\wedge u_1\wedge u_2,& x_{21}(*)=x_{28},\\
x_{22}(*)=x_{3},& x_{23}(*)=\overline{x_{16}},\\
x_{24}(*)=x_{7},& x_{25}(*)=(x_{15}\wedge x_{27}\\
x_{26}(*)=x_{10}\vee\overline{x_{35}},& ~~\wedge x_{34}\wedge x_{37})\\
x_{27}(*)=x_{19},& ~~\vee(x_{27}\wedge x_{31}\wedge x_{34}\wedge x_{37}),\\
x_{28}(*)=x_{29},& x_{29}(*)=x_{12}\vee x_{14},\\
x_{30}(*)=x_{7}\wedge x_{13},&x_{31}(*)=x_{20},\\
x_{32}(*)=x_{8},& x_{33}(*)=x_{24},\\
x_{34}(*)=x_{11},& x_{35}(*)=\overline{x_{4}}\wedge u_3,\\
x_{36}(*)=x_{10}& x_{37}(*)=\overline{x_{4}}\\
~~\vee (x_{20}\wedge x_{35}),& ~~\wedge x_{20}\wedge x_{36}.\\
\end{array}
\end{align}

It is natural to consider the nodes with zero out-degree as observers. Hence we have
\begin{align}\label{6.2}
\begin{array}{l}
y_1=x_1,\\
y_2=x_5,\\
y_3=x_{22},\\
y_4=x_{23},\\
y_5=x_{30}.
\end{array}
\end{align}

As depicted in Fig. \ref{Fig.6.1}, the network is divided into 5 blocks. In the following we use simulation to aggregate them.

\begin{itemize}

\item Consider $\Sigma_1$. Rename the variables as follows:
$$
\begin{array}{lll}
z_1=x_4,& z_2=x_{10},& z_3=x_{20},\\
z_4=x_{26},&z_5=x_{35},&z_6=x_{36},\\
z_7=x_{37};&q_1=x_{20},&q_2=x_{37},
\end{array}
$$
where $q_1$, $q_2$ are block outputs.

Then the block dynamics can be expressed as
\begin{align} \label{6.3}
\begin{cases}
z(t+1)=L_1u(t)z(t),\\
q(t)=H_1z(t),
\end{cases}
\end{align}
where
$$
\begin{array}{l}
L_1=\d_{128}[22,86,22,\cdots,120,56,120]\in {\cal L}_{128\times 1024},\\
H_1=\d_4[1,2,1,\cdots,4,3,4]\in  {\cal L}_{4\times 128}.
\end{array}
$$
Since
$$
H_1L_1(I_8\otimes H_1^T):=T^1
=\begin{bmatrix}
T^1_1&T^1_2\\
\end{bmatrix},
$$
where
\begin{tiny}
$$
T^1_1=\left[
\begin{array}{llllllllllllllll}
 4& 4& 0& 0& 4& 4& 0& 0& 0& 0& 0& 0& 0& 0& 0& 0\\
12&12&16&16&12&12&16&16& 0& 0& 0& 0& 0& 0& 0& 0\\
 4& 4& 0& 0& 4& 4& 0& 0& 8& 8& 0& 0& 8& 8& 0& 0\\
12&12&16&16&12&12&16&16&24&24&32&32&24&24&32&32\\
\end{array}
\right];
$$
$$
T^1_2=\left[
\begin{array}{llllllllllllllll}
 0& 0& 0& 0& 0& 0& 0& 0& 0& 0& 0& 0& 0& 0& 0& 0\\
 0& 0& 0& 0& 0& 0& 0& 0& 0& 0& 0& 0& 0& 0& 0& 0\\
 8& 8& 0& 0& 8& 8& 0& 0& 8& 8& 0& 0& 8& 8& 0& 0\\
24&24&32&32&24&24&32&32&24&24&32&32&24&24&32&32\\
\end{array}
\right].
$$
\end{tiny}
Then the quotient system of (\ref{6.3}) is obtained by updating $T^1$ by
\begin{tiny}
$$
T^1_1=\left[
\begin{array}{llllllllllllllll}
1&1&0&0&1&1&0&0&0&0&0&0&0&0&0&0\\
1&1&1&1&1&1&1&1&0&0&0&0&0&0&0&0\\
1&1&0&0&1&1&0&0&1&1&0&0&1&1&0&0\\
1&1&1&1&1&1&1&1&1&1&1&1&1&1&1&1\\
\end{array}
\right];
$$
$$
T^1_2=\left[
\begin{array}{llllllllllllllll}
 0& 0& 0& 0& 0& 0& 0& 0& 0& 0& 0& 0& 0& 0& 0& 0\\
 0& 0& 0& 0& 0& 0& 0& 0& 0& 0& 0& 0& 0& 0& 0& 0\\
 1& 1& 0& 0& 1& 1& 0& 0& 1& 1& 0& 0& 1& 1& 0& 0\\
 1& 1& 1& 1& 1& 1& 1& 1& 1& 1& 1& 1& 1& 1& 1& 1\\
\end{array}
\right].
$$

\end{tiny}

The probabilistic approximation of (\ref{6.3}) is obtained by updating $T^1$  by
\begin{tiny}
$$
\begin{array}{l}
T^1_1=\\
\left[
\begin{array}{llllllllllllllll}
1/8&1/8&  0&  0&1/8&1/8&  0&  0&  0&  0&0&0&  0&  0&0&0\\
3/8&3/8&1/2&1/2&3/8&3/8&1/2&1/2&  0&  0&0&0&  0&  0&0&0\\
1/8&1/8&  0&  0&1/8&1/8&  0&  0&1/4&1/4&0&0&1/4&1/4&0&0\\
3/8&3/8&1/2&1/2&3/8&3/8&1/2&1/2&3/4&3/4&1&1&3/4&3/4&1&1\\
\end{array}
\right];
\end{array}
$$
$$
T^1_2=\left[
\begin{array}{llllllllllllllll}
  0&  0&0&0&  0&  0&0&0&  0& 0& 0& 0&  0&  0& 0& 0\\
  0&  0&0&0&  0&  0&0&0&  0& 0& 0& 0&  0&  0& 0& 0\\
1/4&1/4&0&0&1/4&1/4&0&0&1/4&1/4&0& 0&1/4&1/4& 0& 0\\
3/4&3/4&1&1&3/4&3/4&1&1&3/4&3/4&1& 1&3/4&3/4& 1& 1\\
\end{array}
\right].
$$
\end{tiny}

\item Consider $\Sigma_2$. Rename the variables as follows:
$$
\begin{array}{lll}
z_1=x_{11},& z_2=x_{15},& z_3=x_{19},\\
z_4=x_{25},&z_5=x_{27},&z_6=x_{31},\\
z_7=x_{34};&v_1=x_{20},&v_2=x_{37};\\
q_1=x_{19},&q_2=x_{25}.&~\\
\end{array}
$$
where $v_1$, $v_2$ are block inputs, $q_1$, $q_2$ are block outputs.

Then the block dynamics can be expressed as
\begin{align} \label{6.4}
\begin{cases}
z(t+1)=L_2v(t)z(t),\\
q(t)=H_2z(t),
\end{cases}
\end{align}
where
$$
\begin{array}{l}
L_2=\d_{128}[1,41,1,\cdots,128,128,128]\in {\cal L}_{128\times 512},\\
H_2=\d_4[1,1,1,\cdots,4,4,4]\in  {\cal L}_{4\times 128}.
\end{array}
$$
Since
$$
H_2L_2(I_4\otimes H_2^T)
:=T^2
$$
where
\begin{tiny}
$$
T^2=\left[
\begin{array}{llllllllllllllll}
 4& 4& 4& 4& 0& 0& 0& 0& 4& 4& 4& 4& 0& 0& 0& 0\\
28&28&28&28& 0& 0& 0& 0&28&28&28&28& 0& 0& 0& 0\\
 0& 0& 0& 0& 0& 0& 0& 0& 0& 0& 0& 0& 0& 0& 0& 0\\
 0& 0& 0& 0&32&32&32&32& 0& 0& 0& 0&32&32&32&32\\
\end{array}
\right].
$$
\end{tiny}

Then the quotient system of (\ref{6.4}) is obtained by updating $T^2$ by
\begin{tiny}
$$
T^2=\left[
\begin{array}{llllllllllllllll}
 1& 1& 1& 1& 0& 0& 0& 0& 1& 1& 1& 1& 0& 0& 0& 0\\
 1& 1& 1& 1& 0& 0& 0& 0& 1& 1& 1& 1& 0& 0& 0& 0\\
 0& 0& 0& 0& 0& 0& 0& 0& 0& 0& 0& 0& 0& 0& 0& 0\\
 0& 0& 0& 0& 1& 1& 1& 1& 0& 0& 0& 0& 1& 1& 1& 1\\
\end{array}
\right].
$$
\end{tiny}

The probabilistic approximation of (\ref{6.4}) is obtained by updating $T^2$  by
\begin{tiny}
$$
\begin{array}{l}
T^2=\\
\left[
\begin{array}{llllllllllllllll}
1/8&1/8&1/8&1/8&0&0&0&0&1/8&1/8&1/8&1/8&0&0&0&0\\
7/8&7/8&7/8&7/8&0&0&0&0&7/8&7/8&7/8&7/8&0&0&0&0\\
  0&  0&  0&  0&0&0&0&0&  0&  0&  0&  0&0&0&0&0\\
  0&  0&  0&  0&1&1&1&1&  0&  0&  0&  0&1&1&1&1\\
\end{array}
\right].
\end{array}
$$
\end{tiny}

\item Consider $\Sigma_3$. Rename the variables as follows:
$$
\begin{array}{lll}
z_1=x_{2},& z_2=x_{3},& z_3=x_{12},\\
z_4=x_{14},&z_5=x_{22},&z_6=x_{29},\\
v_1=x_{19},&v_2=x_{25};&~\\
q_1=y_3=x_{22},&q_2=x_{29}.&~\\
\end{array}
$$
where $v_1$, $v_2$ are block inputs, $q_1=y_3$ is overall system output,  $q_2$ is block output.

Then the block dynamics can be expressed as
\begin{align} \label{6.5}
\begin{cases}
z(t+1)=L_3v(t)z(t),\\
q(t)=H_3z(t),
\end{cases}
\end{align}
where
$$
\begin{array}{l}
L_3=\d_{64}[1,1,1,\cdots,64,64,64]\in {\cal L}_{64\times 256},\\
H_3=\d_4[1,2,3,\cdots,2,3,4]\in  {\cal L}_{4\times 64}.
\end{array}
$$
Since
$$
H_3L_3(I_4\otimes H_3^T)
:=T^3
$$
where
\begin{tiny}
$$
T^3=\left[
\begin{array}{llllllllllllllll}
6&6&6&6&6&6&6&6&6&6&6&6&6&6&6&6\\
2&2&2&2&2&2&2&2&2&2&2&2&2&2&2&2\\
6&6&6&6&6&6&6&6&6&6&6&6&6&6&6&6\\
2&2&2&2&2&2&2&2&2&2&2&2&2&2&2&2\\
\end{array}
\right].
$$
\end{tiny}

Then the quotient system of (\ref{6.5}) is obtained by updating $T^3$ by
\begin{tiny}
$$
T^3=\left[
\begin{array}{llllllllllllllll}
 1& 1& 1& 1& 1& 1& 1& 1& 1& 1& 1& 1& 1& 1& 1& 1\\
 1& 1& 1& 1& 1& 1& 1& 1& 1& 1& 1& 1& 1& 1& 1& 1\\
 1& 1& 1& 1& 1& 1& 1& 1& 1& 1& 1& 1& 1& 1& 1& 1\\
 1& 1& 1& 1& 1& 1& 1& 1& 1& 1& 1& 1& 1& 1& 1& 1\\
\end{array}
\right].
$$
\end{tiny}

The probabilistic approximation of (\ref{6.5}) is obtained by updating $T^3$  by
\begin{tiny}
$$
\begin{array}{l}
T^3=\\
\left[
\begin{array}{llllllllllllllll}
3/8&3/8&3/8&3/8&3/8&3/8&3/8&3/8&3/8&3/8&3/8&3/8&3/8&3/8&3/8&3/8\\
1/8&1/8&1/8&1/8&1/8&1/8&1/8&1/8&1/8&1/8&1/8&1/8&1/8&1/8&1/8&1/8\\
3/8&3/8&3/8&3/8&3/8&3/8&3/8&3/8&3/8&3/8&3/8&3/8&3/8&3/8&3/8&3/8\\
1/8&1/8&1/8&1/8&1/8&1/8&1/8&1/8&1/8&1/8&1/8&1/8&1/8&1/8&1/8&1/8\\
\end{array}
\right].
\end{array}
$$
\end{tiny}

\item Consider $\Sigma_4$. Rename the variables as follows:
$$
\begin{array}{lll}
z_1=x_{7},& z_2=x_{13},& z_3=x_{16},\\
z_4=x_{17},&z_5=x_{23},&z_6=x_{24},\\
z_7=x_{30},&z_8=x_{33};&v_1=x_{25};\\
q_1=y_4=x_{23},&q_2=y_5=x_{30},&q_3=x_{17}.\\
\end{array}
$$
where $v_1$ is a block input, $q_1=y_4$, $q_2=y_5$ are overall system outputs,  $q_3$ is a block output.

Then the block dynamics can be expressed as
\begin{align} \label{6.6}
\begin{cases}
z(t+1)=L_4v(t)z(t),\\
q(t)=H_4z(t),
\end{cases}
\end{align}
where
$$
\begin{array}{l}
L_4=\d_{256}[41,57,41,\cdots,216,200,216]\in {\cal L}_{256\times 512},\\
H_4=\d_8[1,1,5,\cdots,4,8,8]\in  {\cal L}_{8\times 256}.
\end{array}
$$
It is calculated that
$$
H_4L_4(I_2\otimes H_4^T)
:=T^4
$$
where
\begin{tiny}
$$
T^4=\left[
\begin{array}{llllllllllllllll}
2&2&2&2&2&2&2&2&2&2&2&2&2&2&2&2\\
2&2&2&2&2&2&2&2&2&2&2&2&2&2&2&2\\
2&2&2&2&2&2&2&2&2&2&2&2&2&2&2&2\\
2&2&2&2&2&2&2&2&2&2&2&2&2&2&2&2\\
6&6&6&6&6&6&6&6&6&6&6&6&6&6&6&6\\
6&6&6&6&6&6&6&6&6&6&6&6&6&6&6&6\\
6&6&6&6&6&6&6&6&6&6&6&6&6&6&6&6\\
6&6&6&6&6&6&6&6&6&6&6&6&6&6&6&6\\
\end{array}
\right].
$$
\end{tiny}

Then the quotient system of (\ref{6.6}) is obtained by updating $T^4$ by
\begin{tiny}
$$
T^4=\left[
\begin{array}{llllllllllllllll}
 1& 1& 1& 1& 1& 1& 1& 1& 1& 1& 1& 1& 1& 1& 1& 1\\
 1& 1& 1& 1& 1& 1& 1& 1& 1& 1& 1& 1& 1& 1& 1& 1\\
 1& 1& 1& 1& 1& 1& 1& 1& 1& 1& 1& 1& 1& 1& 1& 1\\
 1& 1& 1& 1& 1& 1& 1& 1& 1& 1& 1& 1& 1& 1& 1& 1\\
 1& 1& 1& 1& 1& 1& 1& 1& 1& 1& 1& 1& 1& 1& 1& 1\\
 1& 1& 1& 1& 1& 1& 1& 1& 1& 1& 1& 1& 1& 1& 1& 1\\
 1& 1& 1& 1& 1& 1& 1& 1& 1& 1& 1& 1& 1& 1& 1& 1\\
 1& 1& 1& 1& 1& 1& 1& 1& 1& 1& 1& 1& 1& 1& 1& 1\\
\end{array}
\right].
$$
\end{tiny}

The probabilistic approximation of (\ref{6.6}) is obtained by updating $T^4$  by
\begin{tiny}
$$
T^4=
\left[
\begin{array}{llll}
1/16&1/16&\cdots&1/16\\
1/16&1/16&\cdots&1/16\\
1/16&1/16&\cdots&1/16\\
1/16&1/16&\cdots&1/16\\
3/16&3/16&\cdots&3/16\\
3/16&3/16&\cdots&3/16\\
3/16&3/16&\cdots&3/16\\
3/16&3/16&\cdots&3/16\\
\end{array}
\right]\in {\Upsilon}_{8\times 16}.
$$
\end{tiny}

\item Consider $\Sigma_5$. Rename the variables as follows:
$$
\begin{array}{lll}
z_1=x_{1},& z_2=x_{5},& z_3=x_{6},\\
z_4=x_{8},&z_5=x_{9},&z_6=x_{18},\\
z_7=x_{21},&z_8=x_{28},&z_9=x_{32};\\
v_1=x_{17},&v_2=x_{29};&q_1=y_1=x_{1},\\
q_2=y_2=x_{5}.&~&~\\
\end{array}
$$
where $v_1$, $v_2$ are block inputs, $q_1=y_1$, $q_2=y_2$ are overall system outputs.

Then the block dynamics can be expressed as
\begin{align} \label{6.7}
\begin{cases}
z(t+1)=L_5v(t)z(t),\\
q(t)=H_5z(t),
\end{cases}
\end{align}
where
$$
\begin{array}{l}
L_5=\d_{512}[1,65,5,\cdots,508,448,512]\in {\cal L}_{512\times 2048},\\
H_5=\d_4[1,1,1,\cdots,4,4,4]\in  {\cal L}_{4\times 512}.
\end{array}
$$
It is calculated that
$$
H_5L_5(I_4\otimes H_5^T)
:=T^5
$$
where
\begin{tiny}
$$
T^5=\left[
\begin{array}{llll}
16&16&\cdots&16\\
16&16&\cdots&16\\
48&48&\cdots&48\\
48&48&\cdots&48\\
\end{array}
\right]\in {\cal M}_{4\times 16}.
$$
\end{tiny}

Then the quotient system of (\ref{6.7}) is obtained by updating $T^5$ by
\begin{tiny}
$$
T^5=\J_{4\times 16}
$$
\end{tiny}

The probabilistic approximation of (\ref{6.7}) is obtained by updating $T^5$  by
\begin{tiny}
$$
T^5=
\left[
\begin{array}{llll}
1/8&1/8&\cdots&1/8\\
1/8&1/8&\cdots&1/8\\
3/8&3/8&\cdots&3/8\\
3/8&3/8&\cdots&3/8\\
\end{array}
\right]\in {\Upsilon}_{4\times 16}.
$$
\end{tiny}
\end{itemize}

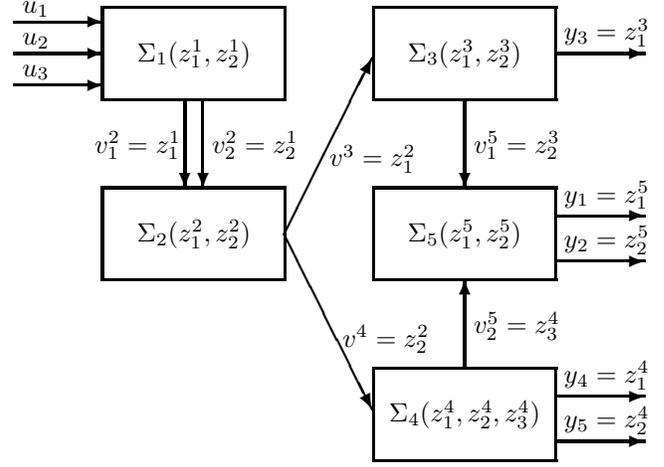
\begin{figure}
\centering
\setlength{\unitlength}{0.6cm}
\begin{picture}(15,11)(-0.5,-0.5)\thicklines
\put(2,8){\framebox(4,2){$\Sigma_1(z^1_1,z^1_2)$}}
\put(2,4){\framebox(4,2){$\Sigma_2(z^2_1,z^2_2)$}}
\put(8,8){\framebox(4,2){$\Sigma_3(z^3_1,z^3_2)$}}
\put(8,4){\framebox(4,2){$\Sigma_5(z^5_1,z^5_2)$}}
\put(8,0){\framebox(4,2){$\Sigma_4(z^4_1,z^4_2,z^4_3)$}}
\put(0,8.3){\vector(1,0){2}}
\put(0,9){\vector(1,0){2}}
\put(0,9.7){\vector(1,0){2}}
\put(0.2,8.5){$u_3$}
\put(0.2,9.2){$u_2$}
\put(0.2,9.9){$u_1$}
\put(3.8,8){\vector(0,-1){2}}
\put(4.2,8){\vector(0,-1){2}}
\put(1.8,6.8){$v^2_1=z^1_1$}
\put(4.4,6.8){$v^2_2=z^1_2$}
\put(6,5){\vector(1,2){1.95}}
\put(7,6.5){$v^3=z^2_1$}
\put(10,8){\vector(0,-1){2}}
\put(10.2,6.8){$v^5_1=z^3_2$}
\put(6,5){\vector(1,-2){1.95}}
\put(7.3,2.5){$v^4=z^2_2$}
\put(10,2){\vector(0,1){2}}
\put(10.2,2.8){$v^5_2=z^4_3$}
\put(12,9){\vector(1,0){2}}
\put(12.2,9.3){$y_3=z^3_1$}
\put(12,5.4){\vector(1,0){2}}
\put(12.2,5.7){$y_1=z^5_1$}
\put(12,4.4){\vector(1,0){2}}
\put(12.2,4.7){$y_2=z^5_2$}
\put(12,1.4){\vector(1,0){2}}
\put(12.2,1.7){$y_4=z^4_1$}
\put(12,0.4){\vector(1,0){2}}
\put(12.2,0.7){$y_5=z^4_2$}

\end{picture}
\caption{Aggregated Simulation of T-Cell \label{Fig.6.2}}
\end{figure}

Summarizing the above blocked simulated systems, an overall aggregated simulation for T-cell is depicted in Fig. \ref{Fig.6.2}, where
$$
\begin{array}{lll}
z^1_1=x_{20},&z^1_2=x_{37},&z^2_1=x_{19},\\
z^2_2=x_{25},&z^3_1=x_{22},&z^3_2=x_{29},\\
z^4_1=x_{23},&z^4_2=x_{30},&z^4_3=x_{17},\\
z^5_1=x_{1},&z^5_2=x_{5}.&~\\
\end{array}
$$

This aggregated simulation of the original system has only $11$ state nodes and $3$ controls. Of course, using the transition form of this quotient system, a further simplification can be done by using $y_i$, $i=1,2,3,4,5$ as new nodes. This is a trad-off between computational load and precision of approximation.

\vskip 5mm

\vskip 5mm

\section{Application to Finite Valued Networks}

\begin{figure}
\centering
\setlength{\unitlength}{0.6cm}
\begin{picture}(6,8)(-0.5,-0.5)\thicklines
\put(0,6){\framebox(5,1){$k$-valued $\Sigma$}}
\put(0,3){\framebox(5,1){Quotient $\Sigma/\sim$}}
\put(0,0){\framebox(5,1){Prob. $\Sigma_P$}}
\put(2.5,6){\vector(0,-1){2}}
\put(2.5,3){\vector(0,-1){2}}
\put(2.5,1){\vector(0,1){2}}
\put(2.7,4.8){Simulation}
\put(2.7,1.8){Bisimulation}
\end{picture}
\caption{Approximation to $k$-valued Networks \label{Fig.7.1}}
\end{figure}
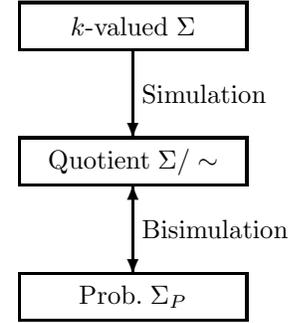

Consider a large scale $k$-valued network $\Sigma$. Assume $\Sigma/\sim$ is its
aggregated simulation and $\Sigma_P$ is its probabilistic form. According to the construction one sees easily that they satisfy the following relation, which is depicted in Fig. \ref{Fig.7.1}.

\begin{prp}\label{p7.1}
\begin{itemize}
\item[(i)] When a $k$-valued (large scale) network $\Sigma$ is approximated by its (aggregated) quotient system, the quotient system is its simulation, denoted by $\Sigma/\sim$.
\item[(ii)] When $\Sigma/\sim$ is approximated by its probabilistic form, $\Sigma_P$, then $(\Sigma/\sim) \approx \Sigma_P$. That is, they are bi-simulated.
\end{itemize}
\end{prp}

\vskip 5mm

\vskip 5mm

Then the relationships (\ref{3.11}) and (\ref{3.12}) imply the following results immediately.

\begin{cor}{ Corollary 7.2}
\begin{itemize}
\item[(i)] Consider Input-Output Decoupling:
$$
\mbox{Solvable for}~~ \Sigma_P~~ \Rightarrow ~~\mbox{Solvable for}~~ \Sigma.
$$
\item[(ii)] Consider Disturbance Decoupling:
$$
\mbox{Solvable for}~~ \Sigma_P ~~\Rightarrow~~ \mbox{Solvable for}~~ \Sigma.
$$
\item[(iii)] Consider Realization:
$$
\mbox{Solvable for}~~ \Sigma_P ~~\Rightarrow ~~\mbox{Solvable for}~~ \Sigma.
$$
\end{itemize}
\end{cor}

If $\Sigma \approx \Sigma_P$ (or $\Sigma \approx \Sigma/\sim$) more properties can be obtained.  We refer to \cite{li18b,jia20,li21,li22} for these.


\begin{thebibliography}{00}

\bibitem{ada03} A. Adamatzky, On dynamically non-trivial three-valued logics: oscillatory and bifurcatory species, {\it Chaos Sokitons Fractals}, Vol. 18, 917-936, 2003.

\bibitem{bel17} C. Belta, B. Yordanov, E.A. Gol, {\it Formal Methods for Discrete-Time Dynamic Systems}, Springer, Switzerland, 2017.
\bibitem{ber68} E. Berlekamp(1968), {\it Algebraic Coding Theory}, McGraw-Hill, New York.
%
\bibitem{bir67} G. Birkhoff(1967), {\it Lattice Theory}, 3 rd ed., Colloq. Pub., Vol. 25, Amer. Math. Soc., Providence.
%
\bibitem{car10}  C. Carlet(2010), {\it Boolean Function for Cryptography and Correcting Codes}, in {\it Boolean Models and Methods in Mathematics}, (Y. Crama, P. Hammer (eds.)), 257-397, Cambridge Univ. Press, Cambridge.
%
\bibitem{che10} D. Cheng, H. Qi, Linea representation of dynamics of Boolean networks, {\it IEEE Trans. Aut. Contr.}, Vol. 55, No. 10, 2251-2258, 2010.
%
\bibitem{che11} D. Cheng, H. Qi, Z. Li (2011), {\it Analysis and Control of Boolean Networks - A Semi-tensor Product
Approach}, Springer, London, 2011.
%
\bibitem{che12}  D. Cheng, H. Qi, Y. Zhao, {\it An Introduction to Semi-tensor Product of Matrices and Its Applications}, World Scientific, Singapo, 2012.
%
\bibitem{che13} T. Chen, U.M. Braga-Neto, Maximum-likelihood estimation of the discrete coefficient of determination if stochastic Noolean systems, {IEEE Trans. Signal Process}, Vol. 61, No. 15, 3880-3894, 2013.
%
\bibitem{che21} D. Cheng, Y. Wu, G. Zhao, S. Fu, A comprehensive survey on STP approach to finite games, {\it J. Sys. Sci. Compl.}, Vol. 34, No. 5, 1666-1680, 2021.
%
\bibitem{che22} D. Cheng, Z. Ji, On networks over finite rings,  {\it J. Franklin Inst.}, Vol. 359, No. 14, 7562-7599, 2022.
%
\bibitem{for16} E. Fornasini, M.E. Valcher, Recent developments in Boolean networks control, {\it J. Contr. Dec.}, Vol. 3, No. 1, 1-18, 2016.
%

\bibitem{ham03} A.G. Hamilton, {\it Logic for Mathematicialn}, Revised Edition, Pub. Tsionhua Univ., Beijing, 2003.
%
\bibitem{hua00} S. Huang, D. Ingber(2000), Shape-dependent control of cell growth, differentiation, and apoptosis: switching between attractors in cell regulatory networks. Exp. Cell Res., Vol. 261, No. 1, 91-103.
%
\bibitem{jipr} Z. Ji, D. Cheng, Control networks over finite lattices, preprint, arxiv:2208.03716, 2022.
%
\bibitem{jia20}  N. Jiang, C. Huang, Y. Chen, J. Kurths, Bisimulation-based stabilization of probabilistic Boolean control networks with state feedback control, {\it Front Inform. Technol. Electron. Eng.}, Vol. 21, No. 2, 268-280, 2020.
%
\bibitem{kau69} S. Kauffman(1969), Metabolic stability and epigenesis in randomly constructed genetic nets, {\it J. Theor. Biol.}, Vol. 22, No. 3, 437-467.
\bibitem{kau93} S. Kauffman(1993), {\it The Origins of Order: Self-organization and Selection in Evolution}, Oxford Univ. Press, London.
%
\bibitem{kau95}  S. Kauffman(1995), {\it At Home in the Universe}, Oxford Univ. Press, London.
%
\bibitem{kla06} S. Klamt, J. Saez-Rodriguez, J.A. Lindquist, L. Simeomi, E.D. Gilles, A methodology for the structural and functional analysis of signaling and regulatory networks, {\it BMC Bioinform.}, Vol. 7, No. 1, Art.no.56, 2006.
%
\bibitem{li18} H. Li, G. Zhao, M. Meng, J. Feng, A survey on applications of semi-tensor product method in engineering, {\it Science China}, Vol. 61, 010202:1-010202:17, 2018.
%
\bibitem{li18b} R. Li, T. Chu, X. Wang, Bisimulations of Boolean control networks, {\it SIAM J. Opt.}, Vol. 51, No. 1, 388-416, 2018.
%
\bibitem{li21} R. Li, Q. Zhang, T. Chu, Reduction and analysis of Boolean control networks by bisimulation, {\it SIAM J. Opt.}, Vol. 59, No. 2, 1033-1056, 2021.
\bibitem{li22} R. Li, Q. Zhang, T. Chu, Bisimulations of Probabilistic Boolean networks, {\it SIAM J. Opt.}, Vol. 60, No. 5, 2631-2657, 2022.
%
\bibitem{lu17}  J. Lu, H. Li, Y. Liu, F. Li, Survey on semi-tensor product method with its applications in logical networks and other finite-valued systems, {\it IET Contr. Thm\& Appl.}, Vol. 11, No. 13, 2040-2047, 2017.
\bibitem{men20} M. Meng, X. Li, G. Xiao, Synchronization of networks over finite fields, {Automatica} 115 (2020) 108877.
%
\bibitem{muh16} A. Muhammad, A. Rushdi, F.A. M. Ghaleb, A tutorial exposition of semi-tensor products of matrices with a stress on their representation of Boolean function, {\it JKAU Comp. Sci.}, Vol. 5, 3-30, 2016.
%
\bibitem{rob86} F. Robert(1986), {\it Discrete Iterations: A Metric Study}, Translated by J. Rolne, Springer, Berlin.
%
\bibitem{sch76} W.M. Schmidt(1976), {\it Equations over Finite Fields, An Elementary Approach}, Springer-Verlag, Berlin.
%
\bibitem{shm02} I. Shmulevich, E. Dougherty, S. Kim, W. Zhang, Probabilistic Boolean networks: a rulebased uncertainty model for gene regulatory networks, {\it Bioinformatics}, Vol. 18, No. 2, 261-274, 2002.
%
\bibitem{tur50}  A.M. Turin(1950), Computing machinery and intelligence,  {\it Mind}, Vol. 59, 433-460. (retitled as ``Can a machine think?" in {\it The world of mathematics}, Vol. 5, Ed. J.R. Newman, et al., New York, 1956).
%
\bibitem{wal92} M. Waldrop, {\it Complexity: The emerging Science at the Edge of Order and Chaos}, Touchstone, New York, 1992.
%
\bibitem{yan22} Y. Yan, D. Cheng, J. Feng, J. Yue(2022), Survey on applications of algebraic state space theory of logical systems to finite state machines, {\it Sci. China Inf. Sci.}, https://doi.org/10.1007/s11432-022-3538-4.
%
\bibitem{yue21} J. Yue, Y. Yan, Z. Chen, H. Deng(2021), State space optimization of finite state machines from the viewpiont of control theory, {\i Front Inform. Technol. Electron. Egg.}, Vol. 22, No. 12, 1598-1609.
%
\bibitem{zho21}  J. Zhong, Y. Pan, D. Lin(2021), On Galois NFSRs equivalent to Fibonacci Ones, In Y. Wu, M. Yung (eds), {\it Information Security and
Cryptology}. Inscript 2020. Lecture Notes in Computer Science (LNCS), Vol. 12612, Springer, Cham., 433-449, 2021.
%
\bibitem{zhu22} S. Zhu, J. Liu, J. Zhong, Y. Liu, J. Cao(2022), Sensors design for large-scale Boolean networks vis pinning observability, {\it IEEE Trans. Aut. Contr.}, Vol. 67, No. 8, 4162-4169.

\end{thebibliography}
\end{document}